\documentstyle[12pt,emulateapj,psfig]{article}

\def\asec{\ifmmode^{\prime\prime}\else$\null^{\prime\prime}$\fi}
\def\deg{\ifmmode^\circ\else$\null^\circ\ $\fi}

\def\etal{{\it et al.\ }}

\def\CO{{\fam0 CO}}
\def\COjone{{\fam0 CO}\ (1-0)}
\def\COjtwo{{\fam0 CO}\ (2-1)}

\def\COtwo{{\fam0 CO_2}}

\def\Ntwo{{\fam0 N_2}}

\def\CHthree{{\fam0 CH_3}}
\def\HCthreeN{{\fam0 HC_3N} }

\def\micron{{\fam0\, \mu m}}

\def\rppr#1#2#3#4#5#6{\rm {#1}\ #2.\ #3.\ {\it #4}\ 
                {\bf #5},\ #6.\par}
\def\rbk#1#2#3#4{\rm { #1}\ #2.\ {\it #3}.\ #4\par}

\begin{document}

\received{ }
\accepted{  }
\journalid{ }{ }
\articleid{ }{ }
\slugcomment{ }

\lefthead{ }
\righthead{ }

\title{CO on Titan: More Evidence for a Well-Mixed 
Vertical Profile}

\altaffiltext{1}{\it phone: (617) 495-7292, FAX: (617) 495-7345, e-mail:
mgurwell@cfa.harvard.edu}

\altaffiltext{2}{\it phone: (626) 395-6186, FAX: (626) 585-1917, e-mail:
dom@gps.caltech.edu}

\center{Mark A. Gurwell\altaffilmark{1}}
\center{\it Harvard--Smithsonian Center for Astrophysics, \\
	60 Garden St., Cambridge, MA 02138}

\centerline{and}

\center{Duane O. Muhleman\altaffilmark{2}}
\center{\it Division of Geological and Planetary Sciences, 
California Institute of Technology, Pasadena, 
California 91125}

\center{Submitted to Icarus: February 9, 2000}

\raggedright

\begin{abstract}

We report new interferometric observations of the $\COjtwo$
rotational transition on Titan.  We find that the spectrum is best fit
by a uniform profile of 52 ppm, with estimated errors of 6 ppm
(40 to 200 km) and 12 ppm (200 to 300 km).

\end{abstract}

\parindent 20pt
\parskip 0pt

\section{Introduction}

The atmosphere of Titan exhibits a complex photochemistry, and many
nitriles and hydrocarbons have been detected by Voyager spacecraft and
from Earth.  Until recently, however, only two oxygen-bearing species
had been detected on Titan: $\COtwo$ (observed by Voyager 1;
\cite{samu1983}) and CO (observed from Earth; \cite{lutz1983}).

The presence of oxygenated molecules is interesting because the
atmosphere of Titan is strongly reducing.  The cold temperatures of
the lower stratosphere and the troposphere imply that $\COtwo$
condenses out of the lower atmosphere and is continuously deposited on
the surface.  To sustain the carbon dioxide abundance a source of
oxygen is needed, and it is generally assumed to be supplied in water
from bombardment of the upper atmosphere by icy grains.  In this model
vaporized water is quickly photolyzed to produce OH, and OH reacts
with hydrocarbon radicals such as $\CHthree$ to produce CO.  CO in
turn reacts with OH to produce $\COtwo$ (\cite{samu1983},
\cite{yung1984}, \cite{toub1995}, \cite{lara1996}). While $\COtwo$ has
a short lifetime (order 10$^3$--10$^4$ years), the photochemical
lifetime of $\CO$ in the atmosphere of Titan is estimated to be very
long ($\sim 10^9$ years; \cite{yung1984}, \cite{chas1991}).

Observationally the missing piece of the oxygen chemistry has been
the source, water.  Recently, water vapor was detected in the
upper atmosphere of Titan by the Short Wavelength Spectrometer (SWS)
aboard the Infrared Satellite Observatory (\cite{cous1998}).  With
observations of the three major components of oxygen chemistry, it is
now possible to check the internal consistency of photochemical
models, and to compare the oxygen chemistry and water infall rate of
Titan with the other giant planets, particularly Saturn
(\cite{feuc1997}, \cite{cous1998}).

Understanding the oxygen chemistry relies on accurate knowledge of the
abundance and distribution of each species.  A longstanding discussion
regarding the CO distribution in Titan's atmosphere, spanning more
than a decade, has been primarily directed toward determining if CO is
well-mixed (\cite{mart1988}, \cite{gurw1995}, \cite{hida1998}).  Since
the residence lifetime of CO is long compared to transport timescales,
the molecular weight of $\CO$ is the same as for the dominant $\Ntwo$
gas, and the atmosphere is never cold enough for CO to condense,
carbon monoxide should be uniformly mixed in the Titan atmosphere to
high altitudes.

Observational data, however, give conflicting results.  Table 1
provides data on the CO abundance as measured by ground-based
observers over the past 17 years.  These observations have been
sensitive to either the troposphere (near- and mid-IR) or the
stratosphere (millimeter).  The data in Table 1 show that no clear
consensus has emerged regarding the CO abundance, either in the
troposphere or the stratosphere.
 
In this Note, we present an analysis of new interferometric
observations of the $\COjtwo$ line on Titan.  The results of this
study have important implications for our understanding of the oxygen
budget and photochemistry of the stratosphere of Titan.

\section{Observations and Data Reduction}

Observations of the $\COjtwo$ rotational transition (rest frequency
$\nu_0$=230.5380 GHz) on Titan were made on November 11 and 12, 1999
with the Owens Valley Radio Observatory Millimeter Array, located near
Big Pine, California.  The Titan ephemeris data was generated using
the Jet Propulsion Laboratory's Horizons on-line system
(\cite{gior1996}).  Titan was approaching eastern elongation with
respect to Saturn, with a separation increasing from $\sim$145$''$ to
more than 200$''$ over the two day period.

The interferometer was aligned in a fairly compact configuration,
providing a synthesized beam of roughly 2\arcsec$\times$2.5\arcsec~ at
the observing frequency, while Titan's apparent surface diameter was
0.86$''$ (at a distance of 8.2165 AU).  Titan was observed on each
night over a period of about 7 hours, when it was above 30$\deg$
elevation.  A single measurement on each baseline consisted of a three
minute integration during which the complex visibility of the source
was recorded.  Amplitude and phase gain variations were monitored
through observations of 0235+164 approximately every 20 minutes, and
antenna pointing was checked about every two hours using 3C84. The
total integration time spent on Titan equaled 238 minutes on each
night.

The signal was detected for each antenna pair in two correlator
systems: a wide-band analog cross correlator ($\sim$1 GHz bandwidth)
and a digital spectrometer.  The CO line is significantly wider than
this system bandwidth, and we utilized two local oscillator tunings to
provide better coverage of the line: on November 11 the digital
spectrometer measured (in two secondary LO tunings) the line in the
upper sideband from $\Delta\nu = -656$ MHz $\rightarrow +32$ MHz, and
on November 12 from $\Delta\nu = -32$ MHz $\rightarrow +656$ MHz.
Spectra in the image sideband, approximately 3 GHz lower in frequency,
were also recorded.  The sideband signals were isolated to better than
20 dB using a phase-switching cycle. The combination of two first and
second LO tunings allowed us to ultimately measure $\pm$650 MHz of the
center frequency of each sideband at 4 MHz resolution, and $\pm$16 MHz
at 0.5 MHz resolution in the line core.  

Calibration of the digital correlator passband was done through
observations of 3C273 and an internal correlated noise source.  The
relative calibration of the sidebands was accurately measured from
observations of Uranus, 3C273, 3C84, and 0235+164 (to $\leq 1$\%),
since nearly all weather and instrumental effects impact each sideband
in a similar manner.  We note that the ability to isolate the
astronomical signal sidebands and record independent spectra in each
sideband represents a significant advantage for interferometric
relative to single-dish observations of CO on Titan because the
emission line is significantly broader than the spectrometer
bandwidth.  In this case the most precise measure of the
line-to-continuum (LTC) ratio is provided by separating the sidebands.
This relative sideband calibration then allows for the production of
an accurate LTC spectrum, with one sideband sensing the continuum, and
the other the line.

The Titan signal strength was sufficient for the application of phase
self-calibration (see \cite{thom1986}) to remove atmospheric phase
variations, which cause decorrelation of the signal, on timescales
shorter than the standard calibration cycle.  After calibration, a
complete spatially unresolved (e.g. ``zero-spacing'') spectrum for
each day was obtained by fitting the observed complex visibilities for
each channel with a model of the Titan visibility function, correcting
for the spatial sampling of the interferometer.  The absolute flux
scale was provided by scaling the continuum sideband intensity to
equal the radiative transfer model flux of Titan at 227.5 GHz,
corrected for the date and time of the observations (see below). The
same scaling factor was applied to the emission line sideband,
preserving the relative calibration.

The resulting combined spectrum of the $\COjtwo$ line on Titan is
shown in Figure 1.  The data clearly shows the $\COjtwo$ line is a
strong emission feature in the spectrum of Titan.  The image sideband
spectrum is essentially flat except for a weak emission line due to
the $\HCthreeN~(25-24)$ rotational transition at 227.419 GHz.  This is
particularly important because it shows that the sideband isolation
procedure was effective to well below the noise level of the spectrum.

\section{Modeling \& Analysis}

The radiative transfer model used to analyze the new CO data is nearly
identical to the one discussed in Gurwell and Muhleman (1995), and we
only highlight important aspects.

The basic parameters of the Titan atmosphere were derived from revised
Voyager 1 radio occultation results (\cite{lind1983}, \cite{lell1990},
\cite{cous1995}), including an atmospheric base at 2575 km from the
center of Titan, with a surface pressure and temperature of 1460
millibar and 96.7 K.  For the thermal profile of the atmosphere we
used an equatorial profile determined by Coustenis and B\'ezard (1995,
their profile A) based upon the occultation results and Voyager 1/IRIS
spectra (Fig.~2) combined with model J of Yelle (1991) for the upper
atmosphere; this same model was used by Hidayat \etal (1998) to
analyze their results.  This model is appropriate since the
observations reported here are unresolved (whole-disk) spectra, which
are heavily weighted by emission from equatorial and low latitudes.

The millimeter continuum opacity on Titan is due to collision induced
dipole absorption by $\Ntwo-\Ntwo$, $\Ntwo-$Ar, and $\Ntwo-$CH$_4$,
and was modeled according to the results of Courtin (1988).  The
spectroscopic parameters for the $\COjtwo$ line were taken from the
JPL catalog (\cite{pick1992}; see also http://spec.jpl.nasa.gov).  The
full Voigt lineshape profile calculation using a fast computational
method (\cite{hui1978}), integrated in pressure over atmospheric
layers of constant temperature, was done using a collisional
line-broadening coefficient for $\COjtwo$ in $\Ntwo$ of $\gamma = 2.21
(T/300 ~ {\fam0 K})^{-0.74}$ MHz mbar$^{-1}$ (\cite{semm1987}).
Radiative transfer calculations at appropriate frequencies were
performed for a variety of radial steps, including limb-sounding
geometries, and integrated over the apparent disk to provide the model
whole-disk spectrum.

The contribution functions ($W(z) = e^{-\tau}d\tau/dz$) for several
frequency offsets from the $\COjtwo$ line center are shown in Fig. 2,
for a single raypath at the disk center. This function describes the
relative contribution of different regions of the atmosphere to the
emitted radiation at each frequency.  The plotted functions assume a
CO abundance of 50 ppm, constant with altitude.  The $\Delta\nu=-3000$
MHz contribution function corresponds to the middle of the continuum
(lower) sideband, and is dominated by the collision induced opacity of
$\Ntwo$.  The other functions correspond to the emission (upper)
sideband, and are dominated by $\COjtwo$ opacity.  The full line
senses the atmosphere from 40 km (the tropopause) to 400 km. However,
the range from 200 to 400 km is sounded mostly in the inner 4 MHz of
the line core.  At 4 MHz spectral resolution we are limited to sensing
the CO abundance from the tropopause to $\sim$200 km.  The 0.5 MHz
spectrum of the line core pushes this upper bound to near 350 km
in the absence of noise.  Thermal noise on the spectral measurements
in practice limit our sensitivity to $\sim 300$ km.

\subsection{Best-fit Uniform CO Distribution}

The radiative transfer model was run for a series of uniform CO
distributions from $q$(CO)=10 to 90 ppm, in steps of 10 ppm.
Resulting spectra are shown in Fig.~1 (in steps of 20 ppm for
clarity).  The model spectra have been convolved to the measurement
spectral resolution of the data in each panel.  The model calculations
show that the continuum (lower) sideband emission is essentially
unaffected by the CO distributions considered, and is an excellent
continuum measurement.  The model gives a flux at 227.5 GHz
of 1.565 Jy for the geometry of the observations, equal to a
disk-average Rayleigh-Jeans brightness temperature of 71.4 K.

Even by eye, the 50 ppm uniform model provides an exceptionally good
fit to the data at both resolutions.  The 50 ppm model gives an RMS
residual of 86.8 mJy for the 4 MHz data, with models of 40 and 60 ppm
giving RMS residuals that are factors of 1.4 and 1.1 times larger,
respectively. Given that a large number of channels are involved
(324), even an 10\% increase in RMS residuals is quite significant.
The 0.5 MHz spectrum is also consistent with this model. A rigorous
least-squares analysis for the best-fit uniform profile gives a formal
solution of 52$\pm$2 ppm from 40 to 300 km.

\subsection{Best-fit Non-Uniform CO Distribution}

An iterative least-squares inversion algorithm (following
\cite{gurw1995}) based on the radiative transfer model was utilized to
solve for a best-fit non-uniform CO distribution.  The logarithm of
the CO distribution was constrained to be a linear function of
altitude.  This constrained solution tests whether a gradient in the
CO distribution is consistent with the observed spectrum.

We find that the best-fit non-uniform profile, with formal error, is
48$\pm$4 ppm at 40 km, rising to 60$\pm$10 ppm at 300 km.  The RMS
residual is 86.1 mJy, representing less than 1\% improvement in the
residual over the best-fit uniform profile.

\subsection{Error Estimates and the Best-fit CO Distribution}

The formal errors quoted in the above sections are the direct results
of the least-squares analyses, and therefore do not take into account
errors in the radiative transfer modeling or the calibration of the
spectrum.

To test whether the continuum emission model is a serious source of
error (since the spectrum is calibrated by referencing to the
continuum sideband data), we recomputed the continuum emission at
227.5 GHz, scaling the collision induced dipole absorption calculated
from the data of Courtin (1988) by factors of 0.5 and 2.  The
calculated emission results were indistinguishable from the nominal
model, which can be explained as the result of two factors.  First,
the collision induced continuum absorption scales as the square of
pressure, and is therefore a very steep function of
altitude. Therefore, increasing (or decreasing) the absorption
coefficient even by factors of two will only increase the peak of the
contribution function by a small fraction of a scale height. Second,
the peak of the contribution function is right at the tropopause,
where the temperature gradient is near zero.  The result is that the
emission change is very small, and we estimate that this error is
about 1\%.  

The spectrum sidebands were calibrated assuming the QSO calibration
sources had a spectral index of -0.5 (e.g. flux $\propto \nu^{-0.5}$).
However, the spectral index of these types of sources vary over the
range of 0 to -1, and could lead to a calibration error of
approximately 1\% in the relative calibration over the 3 GHz
difference in the sidebands.

Adding the calibration errors in quadrature, we find an error in the
relatively calibrated spectrum of about 1.4\%.  Using the uniform
distribution models discussed in section 3.1, we find that a 3\% error
in the relative calibration could lead to an error of roughly 10 ppm
in a worst-case situation (we note that this does not include a
refitting of the {\it lineshape}, which would tend to reduce this
error; hence this is a worst-case estimate).  The calibration error is
then about 6 ppm using this scaling.  For the uniform model solution,
the formal error is significantly smaller than this calibration error
estimate, and we believe that the error of our measurement is
therefore about 6 ppm.

Noting that the non-uniform solution only improves the RMS residuals by
1\% at best, and that the formal errors on the non-uniform solution
encompass our uniform solution, we favor the uniform model for the CO
distribution, which is in agreement with the current understanding of
the chemistry of CO in the atmosphere of Titan.  The high resolution data
provides the information on altitudes above 200 km, and as can be seen
in Fig.~1 this data has a higher RMS noise (by a factor of $\sim$2);
this increases our error estimate by a factor of about two over this
altitude range. We therefore find that the $\COjtwo$ spectrum is best fit
by a uniform profile of 52 ppm, with estimated errors of 6 ppm
(40 to 200 km) and 12 ppm (200 to 300 km).

\section{Discussion}

The results presented here are nearly identical to our previous
estimate of the CO distribution based on observations of the $\COjone$
transition (\cite{gurw1995}) and consistent with the original
measurement of tropospheric CO (\cite{lutz1983}).  Taken together,
these measurements suggest a vertical profile of CO that is constant
with altitude, at about 52 ppm, from the surface to at least 300 km.

These results are at odds with the recent measurements of Noll (1996),
who found a tropospheric abundance of 10 ppm, and Hidayat \etal
(1998), who found a stratospheric CO abundance of around 27 ppm (Table
1).  Noll (1996) explored the possibility that their simple reflecting
layer was not the surface, but a higher altitude 'haze' layer.  If the
reflecting layer was at 0.9 bar (14 km) the spectrum was best fit with
a CO abundance of 60 ppm.  However, based on other evidence they found
this model less satisfactory than a surface reflecting layer.  The
results of Hidayat \etal come from an analysis of several lines of CO,
including the $\COjone$ and $\COjtwo$ lines; the discrepancy between
their results and ours does not appear to be due to differences in
modeling the atmosphere of Titan, but derives from differences in the
measurement techniques and the resulting calibrated spectra
(A.~Marten, personal communication).  However, we do point out that
the interferometric method does offer advantages over single-dish
observations for measuring the very broad lines of CO from the
atmosphere of Titan.

We find the model of a uniform distribution of CO in the atmosphere of
Titan provides a good fit to our data, but we cannot rule out a
difference between the tropospheric and stratospheric CO abundance,
since our data is insensitive to the lower atmosphere.  A final
confirmation of the abundance of CO and its vertical distribution
requires further near- and mid-IR measurements of CO in the
troposphere.

\vskip 25pt

\center{\bf Acknowledgements}

This work was supported in part by NASA grant NAG5-7946.

\vfill

\centerline{TABLE 1. Observations of CO in Titan's Atmosphere}
\label{tab:coobs}
\begin{center}
\begin{tabular}{cccl}
\hline\noalign{\smallskip}

Altitude & Mixing ratio (ppm)$^a$ & Wavelength & ~~~Reference \\

\noalign{\smallskip}\hline\noalign{\smallskip}

Troposphere  & 48$^{+100}_{-32}$ & 1.57$\micron$    & \cite{lutz1983} \\
Stratosphere & 60$\pm$40$^b$     & 2.6 mm           & \cite{muhl1984} \\
Stratosphere & 2$^{+2}_{-1}$     & 2.6 mm           & \cite{mart1988} \\
Stratosphere & 50$\pm$10         & 2.6 mm           & \cite{gurw1995} \\
Troposphere  & 10$^{+10}_{-5}$   & 4.8$\micron$     & \cite{noll1996} \\
Stratosphere & 27$\pm$5$^c$      & 2.6, 1.3, 0.9 mm & \cite{hida1998} \\

\noalign{\smallskip}\hline\noalign{\smallskip}

Stratosphere & 52$\pm$6          & 1.3 mm           & this work \\

\noalign{\smallskip}\hline\noalign{\smallskip}
\noalign{\smallskip}

\multispan4 {$^a$Mixing ratio defined as N(CO)/N(Total), i.e.~{\it
 not} referenced to $\Ntwo$. \hfill} \\

\noalign{\smallskip}

\multispan4 {$^b$Reanalyzed by Paubert \etal (1984): 75$^{+105}_{-45}$
ppm \hfill}\\

\noalign{\smallskip}

\multispan4 {$^c$Non-uniform model: 29$\pm$5 ppm (60 km), 24$\pm$5
ppm (175 km), 4.8$\pm$2 ppm (350 km) \hfill}\\

\end{tabular}
\end{center}


\vfill

\begin{center}
{\sc Figure Captions}
\end{center}
\figcaption{ Calibrated $\COjtwo$ spectrum obtained November 11--12,
1999 with the OVRO millimeter interferometer, and model spectra
calculated assuming uniform profiles of $q$(CO)=10, 30, 50, 70, and 90
ppm.  The largest panel shows the full spectrum in both sidebands at 4
MHz resolution. The continuum sideband is flat except for a weak,
narrow emission feature due to the $\HCthreeN (25-24)$ rotational
transition near $\Delta\nu=-3119$ MHz. The larger inset concentrates
on the upper sideband data, containing the CO emission feature
($\pm$650 MHz of the line center) at 4 MHz resolution.  The small
inset provides the inner $\pm$16 MHz of the line core at 0.5 MHz
resolution.  For each panel, the model spectra are convolved to the
spectral resolution of the data.
\label{fig1}}

\figcaption{({\it left}) Contribution functions for selected frequencies
near the $\COjtwo$ line center for the Titan atmosphere, at normal
incidence. These functions were calculated assuming a uniform 
profile $q$(CO)=50 ppm. ~ ({\it right}) The model atmospheric
temperature profile derived from Voyager 1 radio occultation
measurements and IRIS spectra, and smoothly merged to fit a non-LTE
model of the upper atmosphere (adopted from \cite{cous1995}).
\label{fig2} }

\vfill

\newpage

\centerline{\psfig{figure=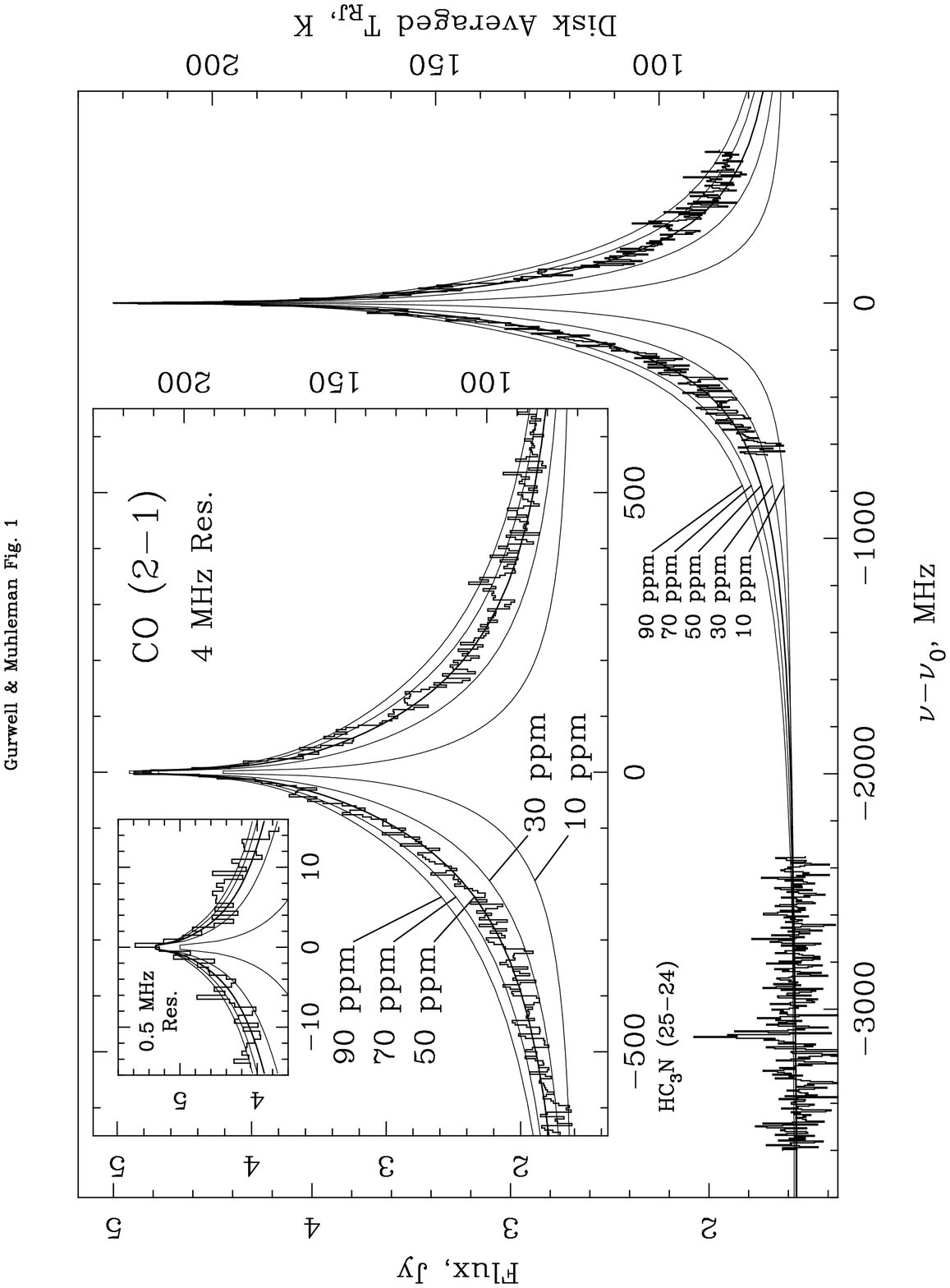,angle=0.,height=\vsize}}

\newpage

\centerline{\psfig{figure=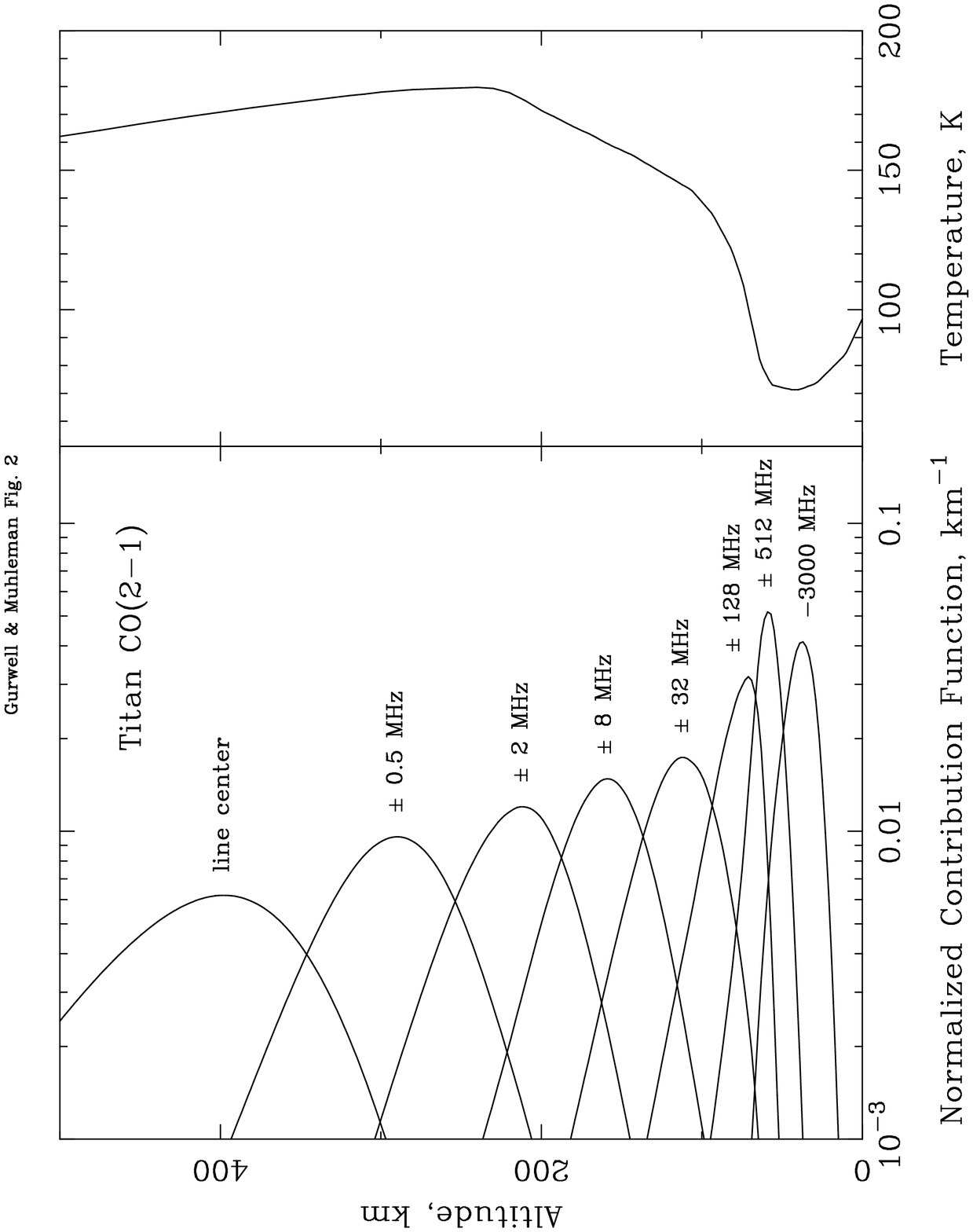,angle=0.,height=\vsize}}
			     

\begin{thebibliography}{ }



\bibitem[Chassefi\`ere and Cabane 1991]{chas1991}\rppr{
Chassefi\`ere, D.~and M.~Cabane}{1991}{Stratospheric depletion of CO
on Titan} {Geophys.~Res.~Lett.}{18}{467-470}

\bibitem[Courtin 1988]{cour1988}\rppr{Courtin, R.}{1988}{Pressure-induced
absorption coefficients for radiative transfer calculations in Titan's
atmosphere}{Icarus}{75}{245-254}
 
\bibitem[Coustenis and B\'ezard 1995]{cous1995}\rppr{Coustenis, A.,
and B.~B\'ezard}{1995}{Titan's atmosphere from Voyager infrared
observations ~ IV. Latitudinal variations of temperature and
composition}{Icarus}{115}{126-140}

\bibitem[Coustenis \etal 1998]{cous1998}\rppr{Coustenis, A.,
A.~Salama, E.~Lellouch, Th.~Encrenaz, G.L.~Bjoraker, S.E.~Samuelson,
Th.~de Graauw, H.~Feuchtgruber, and M.F.~Kessler}{1998}{Evidence for
water vapor in Titan's atmosphere from ISO/SWS
data}{Astron.~Astrophys.}{336}{L85-L89}

\bibitem[Feuchtgruber \etal 1997]{feuc1997}\rppr{Feuchtgruber, H.,
E.~Lellouch, Th.~de Graauw, B.~B\'ezard, Th.~Encrenaz, and
M.~Griffin}{1997}{External supply of oxygen tot he atmospheres of the
giant planets}{Nature}{389}{159-162}

\bibitem[Giorgini \etal 1996]{gior1996}\rppr{Giorgini, J.D.,
D.K.~Yeomans, A.B.~Chamberlin, P.W.~Chodas, R.A.~Jacobson,
M.S.~Keesey, J.H.~Lieske, S.J.~Ostro, E.M.~ Standish, and
R.N.~Wimberly}{1996}{JPL's on-line solar system data
service}{B.A.A.S.}{28}{No.~3, 1158}

\bibitem[Gurwell and Muhleman 1995]{gurw1995}\rppr{Gurwell, M.A., and
D.O.~Muhleman}{1995}{CO on Titan: Evidence for a well-mixed vertical
profile}{Icarus}{117}{375-382}

\bibitem[Hidayat \etal 1998]{hida1998}\rppr{Hidayat, T., A.~Marten,
B.~B\'ezard, D.~Gautier, T.~Owen, H.E.~Matthews, and G.~Paubert}
{1998}{Millimeter and submillimeter heterodyne observations of Titan:
the vertical profile of carbon monoxide in its stratosphere}
{Icarus}{133}{109-133}

\bibitem[Hui \etal 1978]{hui1978}\rppr{Hui, A.K., B.H.~Armstrong, and
A.A. Wray}{1978}{Rapid computation of the Voigt and complex error
functions}{J.~Quant.~Spectrosc.~Radiat.~Transfer}{19}{509-516}
 
\bibitem[Lara \etal 1996]{lara1996}\rppr{Lara, L.M., E.~Lellouch,
J.J.~L\'opez-Moreno, and R.~Rodrigo}{1996}{Vertical distribution of
Titan's atmospheric neutral constituents}{J.~Geophys.~Res.}{101}
{23261-23238}

\bibitem[Lellouch 1990]{lell1990}\rppr{Lellouch, E.}{1990}{Atmospheric
models of Titan and Triton}{Ann.~Geophysicae}{8}{653-660}
 
\bibitem[Lindal \etal 1983]{lind1983}\rppr{Lindal, G.F., G.E.~Wood,
H.B.~Hotz, D.N.~Sweetnam, V.R.~Eshleman, and G.L.~Tyler}{1983}{The
atmosphere of Titan: An analysis of the Voyager 1 radio occultation
measurements}{Icarus}{53}{348-363}
 
\bibitem[Lutz \etal 1983]{lutz1983}\rppr{Lutz, B.L., C.~de Bergh,
and T.~Owen}{1983}{Titan: Discovery of carbon monoxide in its
atmosphere}{Science}{220}{1374-1375}
 
\bibitem[Marten \etal 1988]{mart1988}\rppr{Marten, A., D.~Gautier,
L.~Tanguy, A.~Lecacheux, C.~Rosolen, and G.~Paubert}{1988}{Abundance
of carbon monoxide in the stratosphere of Titan from millimeter
heterodyne observations}{Icarus}{76}{558-562}
 
\bibitem[Muhleman \etal 1984]{muhl1984}\rppr{Muhleman,
D.O., G.L.~Berge, and R.T.~Clancy}{1984}{Microwave measurements of
carbon monoxide on Titan}{Science}{223}{393-396}
 
\bibitem[Noll \etal 1996]{noll1996}\rppr{Noll, K.S., T.R.~Geballe,
R.F.~Knacke, and Y.J.~Pendleton}{1996}{Titan's 5$\micron$ spectral
window: carbon monoxide and the albedo of the surface}
{Icarus}{124}{625-631}

\bibitem[Paubert \etal 1984]{paub1994}\rppr{Paubert, G., D.~Gautier,
and R.~Courtin}{1984}{The millimeter spectrum of Titan: Detectability
of HCN, HC$_3$N, CH$_3$CN, and the CO abundance.}{Icarus}{60}{599-612}
 
\bibitem[Pickett \etal 1992]{pick1992}\rbk{Pickett, H.M.,
R.L.~Poynter, and E.A.~Cohen}{1992} {Submillimeter, Millimeter and
Microwave Spectral Line Catalog}{JPL Publication 80--23, Rev. 3}
 
\bibitem[Samuelson \etal 1983]{samu1983}\rppr{Samuelson, R.E.,
W.C.~Maguire, R.A.~Hanel, V.G.~Kunde, D.E.~Jennings, Y.L.~Yung,
and A.C.~Aikin}{1983}{$\COtwo$ on Titan} {J.~Geophys.~Res.}
{88}{8709-8715}

\bibitem[Semmoud-Monnanteuil and Colmont 1987]{semm1987}\rppr{
Semmoud-Monnanteuil, N., and J.M.~Colmont}{1987}{Pressure broadening of
millimeter lines of carbon monoxide}{J.~Molec.~Spec.}{126}{210-219}

\bibitem[Thompson, Moran, and Swenson 1986]{thom1986}\rbk{Thompson,
A.R., J.M.~Moran, and G.W.~Swenson Jr.}{1986}{Interferometry and
Synthesis in Radio Astronomy}{1st ed. Wiley, New York}

\bibitem[Toublanc \etal 1995]{toub1995}\rppr{Toublanc, D.,
J.P.~Parisot, J.~Brillet, D.~Gautier, F.~Raulin, and
C.P.~McKay}{1995}{Photochemical modeling of Titan's
atmosphere}{Icarus}{113}{2-26}

\bibitem[Yelle 1991]{yell1991}\rppr{Yelle, R.V.}{1991}{Non-LTE models
of Titan's upper atmosphere}{Astrophys.~J.}{383}{380-400}
 
\bibitem[Yung \etal 1984]{yung1984}\rppr{Yung, Y.L., M.~Allen, and
J.P.~Pinto}{1984}{Photochemistry of the atmosphere of Titan:
Comparison between model and observations}
{Astrophys.~J.~Suppl.~Ser.}{55}{465-506}

\end{thebibliography}
\end{document}